\documentclass[smallextended]{svjour3}   

\usepackage{graphicx}                                   
\usepackage{mathptmx}                                   
\usepackage{amsmath}                                            
\usepackage{amsfonts}                                  
\usepackage{amssymb}                                            
\usepackage[caption=false]{subfig}
\usepackage{marvosym}
\usepackage[top=2.5cm,bottom=2.5cm,margin=2.5cm]{geometry}

\usepackage{ifpdf}
\ifpdf
\usepackage[unicode,colorlinks,linkcolor=blue,urlcolor=blue,citecolor=blue,
plainpages=false,pdfpagelabels,breaklinks]{hyperref}
\else
\usepackage[unicode,colorlinks,linkcolor=blue,urlcolor=blue,citecolor=blue,
plainpages=false,pdfpagelabels,linktocpage]{hyperref}
\fi

\usepackage[numbers,square,sort&compress]{natbib}

\usepackage{color}

\begin{document}

\title{On the accuracy of the IWM--CFC approximation in differentially rotating relativistic stars}

\titlerunning{Accuracy of the IWM--CFC approximation} 

\author{Panagiotis Iosif \and Nikolaos Stergioulas}

\authorrunning{P. Iosif \and N. Stergioulas} 

\institute{P. Iosif \and N. Stergioulas \at
              Department of Physics, Aristotle University of Thessaloniki, Thessaloniki 54124, Greece \\
              \email{piosif@auth.gr , niksterg@auth.gr} 
}

\date{Received: date / Accepted: date}
\maketitle      
  
\begin{abstract}
We determine the accuracy of the conformal flatness (IWM--CFC) approximation for the case of single, but strongly differentially rotating relativistic stars. We find that for the fastest rotating and most relativistic polytropic models, the deviation from full general relativity is below 5\% for integrated quantities and below 10\% for local quantities, such as the angular velocity. Furthermore, we study the deviation of the IWM--CFC approximation from full general relativity by evaluating and comparing different error indicators. We find that for the models that are not near the maximum mass, a simple error indicator constructed from local values of the metric potentials is more indicative of the accuracy of the IWM--CFC approximation than an error indicator that is based on the Cotton--York tensor. Furthermore, we construct a simple, linear empirical relation that allows for the estimation of the error made by the IWM--CFC approximation and which only involves  the flattening of the star  due to rotation and the minimum value of the lapse function. Thus, in any numerical simulation involving rotating relativistic stars, one can readily know the deviations from full general relativity due to the IWM--CFC approximation.
\end{abstract}

\section{Introduction}
\label{sec:intro}
Binary neutron star mergers can have different outcomes, depending on the  neutron star masses and on the equation of state of high density matter. One possible outcome of such a merger is a long-lived ($\tau > 10 \, \mathrm{ms} $) compact remnant, a hypermassive neutron star (HMNS) \cite{Baumgarte2000}, supported by strong differential rotation e.g. \cite{Hotokezaka2011, Bauswein2012}. Recently, a triplet of oscillation frequencies of the merger remnant has been identified in the gravitational waves emitted in the post-merger phase, in simulations \cite{NS_Bauswein_Zagkouris_Janka_2011} employing the spatially-conformally-flat spacetime approximation (IWM--CFC \cite{isenberg7808, wilsonCFC96}). Evaluating the accuracy of the IWM--CFC approximation in different astrophysically relevant systems is crucial in order to better understand the limits within which it can be applied. Here, we study the accuracy of the IWM--CFC approximation in stars that are highly deformed due to strong differential rotation, as it is the case for HMNSs.

The IWM--CFC approximation has already been tested for the case of single, uniformly rotating, relativistic stars \cite{CST96}. Furthermore, it has been applied in the study of gravitational collapse \cite{Dimmelmeier2002a, Dimmelmeier2002b, Saijo2004, ott07_CFC_acc} where the deviation from general relativity was shown to be negligible \cite{ShibataSekiguchi2004}. In addition, it was employed in studying neutron star oscillations and stability \cite{Dimmelmeier2006, CerdaDuran2008, Abdikamalov2009} as well as for the construction of binary neutron star initial data (see \cite{Miller_Gressman_Suen} for an evaluation of the accuracy in this case) and in the simulation of their mergers (e.g. \cite{Oechslin2006, Oechslin2007, NS_Bauswein_Zagkouris_Janka_2011, Bauswein2014}). In \cite{KleySchafer1999} it was shown that the IWM--CFC approximation has similar accuracy to the first post-Newtonian approximation for the case of rigidly rotating, relativistic disks of dust. Extended versions of IWM--CFC \cite{Cordero2009} were also implemented in other computational codes, e.g. \cite{CerdaDuran2005, Bucciantini2011}.

 Here, we focus on differentially rotating polytropes with polytropic index in the range of $N=0.5-1.0 $ and find that for the fastest rotating and most relativistic models, the deviation from full general relativity is below 5\% for integrated quantities and below  10\% for local quantities, such as the angular velocity. We construct different error indicators and find that a simple error indicator constructed from local values of the metric potentials is superior compared to another one based on the Cotton--York tensor. Finally, we construct a simple, linear empirical relation that allows for the estimation of the error made by the IWM--CFC approximation and which only involves  the flattening of the star due to rotation and the minimum value of the lapse function. This empirical relation will be useful in evaluating the accuracy of numerical simulations. 

The paper is organized as follows: in Section \ref{sec:method} we present the main equations that govern the models we construct. In Section \ref{sec:diagnostics} we provide details for the structure of the equilibrium models and for the different diagnostics we construct, while Section \ref{sec:results} contains our numerical results. In Section \ref{sec:discussion} we compare the different diagnostics and conclude.
If otherwise not specified, non-dimensional units are enforced by the three conditions $c=G=M_\odot=1$. 

\section{Main equations and method} 
\label{sec:method}
In full general relativity (GR), the line element for stationary, axisymmetric stars in equilibrium is given by (see e.g. \cite{FriedmanStergioulas2013})
\begin{equation}
ds^2 = -e^{\gamma + \rho} dt^2 + e^{\gamma - \rho} r^2 \sin^2 
\theta (d\phi - \omega dt)^2 + e^{2\mu} (dr^2 + r^2 d\theta^2) \label{eq:KEH_metric} \; ,
\end{equation}
where $ \gamma $, $ \rho $, $ \omega $ and $ \mu $ are metric potentials depending only on the coordinates $ r $ and $ \theta $. We model the stellar matter as a perfect fluid and assume that the equation of state (EOS) is described by the polytropic relation
\begin{equation}
p = K \rho ^{1 + \frac{1}{N}} \label{eq:polytropic_EOS} \; ,
\end{equation}
where $ \rho $ is the rest mass density, $ K $ the polytropic constant and $ N $ the polytropic index. Following \cite{KEH89a,KEH89b}, we adopt a simple differential rotation law of the form
\begin{equation}
u^t u_\phi := F(\Omega) = A^2 (\Omega_c - \Omega) \label{eq:rotation_law} \; ,
\end{equation}
where $ A $ is a positive constant that determines the length scale over which the angular velocity $\Omega$ changes within the star and $ \Omega_c $ is the angular velocity at the center of the configuration.
 
In the $3+1$ formalism of general relativity, the line element is written as
\begin{equation}
ds^2 = -\alpha^2 dt^2 + \gamma_{ij} (dx^i + \beta^i dt)(dx^j + \beta^j dt),
\end{equation}
where $\alpha$ is the lapse function, $\beta^i$ is the shift vector and $\gamma_{ij}$ is the spatial metric. In the context of IWM--CFC, we assume that the spatial metric is conformally flat, i.e.
\begin{equation}
\gamma_{ij} = \psi^4 \eta_{ij}, 
\end{equation}
where $\psi$ is a conformal factor and $\eta_{ij}$ is the flat metric. In spherical-like coordinates, for an axisymmetric star (and in the absence of meridional circulation) $\beta^{\phi} $ is the only non-zero component of the shift vector  $ \beta^{i} $. Thus, the line element in the IWM--CFC approximation and $3+1$ split is written as 
\begin{equation}
ds^2 = -\alpha^2 dt^2 + \psi^4 (dr^2 + r^2 d \theta^2) +  \psi^4 r^2 \sin ^2 \theta (d \phi + \beta^{\phi} dt)^2  \label{eq:CFC_3+1_metric} \; .
\end{equation}
Inspection of the line elements (\ref{eq:KEH_metric}) and (\ref{eq:CFC_3+1_metric}) yields the relations
\begin{equation}
\alpha = e ^ {(\gamma + \rho) / 2} \; , \qquad
\psi = e ^ {\mu / 2 } = e ^{ (\gamma - \rho) / 4} \; , \qquad 
\beta^{\phi} = -\omega \label{eq:lapse_shift_psi} \; ,
\end{equation}
and therefore metric (\ref{eq:KEH_metric}) takes the IWM--CFC form, if
\begin{equation}
\mu = \frac{\gamma - \rho}{2} \label{eq:mu_imposeCFC} \; .
\end{equation}

In practice, one can easily convert a code that solves for the full metric (\ref{eq:KEH_metric}) into a code that implements the IWM--CFC approximation. Instead of solving for the metric potential $\mu$, one can impose the condition (\ref{eq:mu_imposeCFC}) between the three metric functions, thus allowing for the construction of numerical models in the IWM--CFC approximation without any other modification in a computational code that implements, for example, the Komatsu, Eriguchi and Hachisu (KEH) scheme \cite{KEH89a,KEH89b}. Such a code is {\tt rns} \cite{RNScode, stergfried95, nozawa98}, a new version of which was used as a basis for implementing the condition (\ref{eq:mu_imposeCFC}).

\section{Equilibrium models and diagnostics}
\label{sec:diagnostics}
In order to test the accuracy of the IWM--CFC approximation, we initially focus on the two sequences (A and B) of differentially rotating models with $ N=1 $ and $ K=100 $, originally presented in Table 1 of \cite{SAF2004}. Models of sequence A have a constant rest mass of $ M_0 = 1.506 $, while models of sequence B have a constant central energy density of $ \epsilon_c = 1.444 \times 10^{-3} $ (or equivalently a constant central mass density $ \rho_c = 1.28 \times 10^{-3} $). For comparison, the central energy density and rest mass density for the maximum mass nonrotating model are $ \epsilon_c = 3.9 \times 10^{-3} $ and $ \rho_c = 3 \times 10^{-3} $, having a gravitational mass of $ M = 1.636 $ and a rest mass of $ M_0 = 1.797 $. The value of $ M_0 $ used in sequence A is chosen in order to represent a nascent, differentially rotating neutron star. In addition to the models presented in \cite{SAF2004}, we constructed two models that have even higher rotation rates: model A12, with a polar to equatorial axis ratio of $ r_p / r_e =0.25$ and model B13 with $ r_p / r_e = 0.34 $.

In order to study the most relativistic stable  models for this EOS,  we construct an additional differentially rotating sequence C. Taking into account \cite{Giacomazzo2011, Takami2011} we fix the central energy density of sequence C at $ \epsilon_c = 3.3 \times 10^{-3} $ (equivalently the central rest-mass density is $ \rho_c = 2.6 \times 10^{-3} $). Table \ref{tab:seqC_characteristics} summarizes the characteristics of this new sequence. We note that model C0 has a compactness of $ M/R = 0.2 $, while models A0 and B0 (which coincide since they have the same central energy density $ \epsilon_c = 1.444 \times 10^{-3} $, see \cite{SAF2004} for details), have a compactness of 0.15.

\begin{table}
\centering
\caption{\label{tab:seqC_characteristics}Characteristics of the differentially rotating sequence C. A polytropic equation of state is assumed with $ N=1 $ and $ K=100 $. The central energy density is fixed at $ 3.3 \times 10^{-3} $ (which corresponds to a  central rest-mass density of $ \rho_c = 2.6 \times 10^{-3} $). $ M $ is the gravitational mass, $ T /|W| $ is the ratio of the rotational over the gravitational binding energy, $ \Omega_c $ is the angular velocity at the center of the configuration, $\Omega_e$ is the angular velocity at the equator and $R_e$ is the circumferential radius at the equator (all quantities are for nondimensional units set by $c=G=M_\odot=1$). }
\begin{tabular}{@{}llllllll}
\hline\noalign{\smallskip}
Model & $ \epsilon_c $ & $ r_p / r_e $ & $ M $ & $ T /|W| $ & $ \Omega_c $ & $\Omega_e$ & $R_e$ \\
    & ($ \times 10^{-3}$) & & & ($ \times 10^{-1}$) & ($ \times 10^{-2}$) & ($ \times 10^{-2}$) & \\
\noalign{\smallskip}\hline\noalign{\smallskip}
C0 & 3.3 & 1.0 & 1.626 & 0.000 & 0.000 & 0.000 & 8.07\\
C1 & 3.3 & 0.9 & 1.692 & 0.251 & 4.114 & 1.293 & 8.26\\
C2 & 3.3 & 0.8 & 1.774 & 0.534 & 6.053 & 1.854 & 8.47\\
C3 & 3.3 & 0.7 & 1.879 & 0.851 & 7.798 & 2.307 & 8.69\\
C4 & 3.3 & 0.6 & 2.011 & 1.205 & 9.671 & 2.723 & 8.89\\
C5 & 3.3 & 0.5 & 2.167 & 1.587 & 12.11 & 3.169 & 8.95\\
C6 & 3.3 & 0.4 & 2.298 & 1.963 & 16.05 & 3.759 & 8.68\\
C7 & 3.3 & 0.3 & 2.306 & 2.285 & 22.59 & 4.614 & 8.00\\
\noalign{\smallskip}\hline
\end{tabular}
\end{table}

For each model that is constructed, various physical quantities are calculated both in IWM--CFC and in full GR. As a first diagnostic, we evaluate relative differences between the two approaches for the gravitational mass $M$, the ratio $ T / |W| $ of the rotational kinetic energy over the gravitational binding energy, the circumferential radius $R_e$ at the equator and the angular velocity $\Omega_c$ at the center of the configuration.

Furthermore, we calculate two more measures of the accuracy of IWM--CFC. Following \cite{CST96}, a simple error indicator is given by the expression
\begin{equation}
\Delta_{1} := \mathrm{max} \left| \frac{\mu - \bar{\mu} }{\mu} \right|
\label{eq:Delta1} \; ,
\end{equation}
where $ \bar {\mu} := (\gamma - \rho)/2 $, i.e. the term appearing in the flatness condition (\ref{eq:mu_imposeCFC}). The above expression vanishes in IWM--CFC, but is non-vanishing for models constructed in full GR with the metric (\ref{eq:KEH_metric}). The maximum value is taken over the whole grid. We note that the metric potentials appearing in (\ref{eq:Delta1}) are calculated in full GR and as a result, $\Delta_{1}$ can be calculated in one run of the code. For comparison and since $ \mu $ is involved in the calculation of $\gamma$ and $\rho$, we also define a variant of the previous diagnostic given by the expression
\begin{equation}
\Delta_{2} := \mathrm{max} \left| \frac{\mu_{\mathrm{GR}}-\mu_{\mathrm{CFC}}}{\mu_{\mathrm{CFC}}}  \right| \label{eq:Delta2} \; ,  
\end{equation}
where two separate computations (one in full GR and another in IWM--CFC) are required (the central density and axis ratio are kept fixed).

Another possible diagnostic, proposed in \cite{Miller_Gressman_Suen}, is constructed by considering the Cotton--York tensor \cite{York1971} defined on a 3-dimensional spacelike hypersurface
\begin{equation}
C^{ij} := 
\epsilon^{ik\ell} D_k \left(R^j _{\;\ell} - \frac{1}{4} R \delta^j _{\;\ell} \right) \label{eq:CY_tensor} \; ,
\end{equation}
where $ D_i $ is the 3-dimensional covariant derivative associated with the spatial metric $ \gamma_{ij} $, $ R_{ij} $ is the 3-dimensional Ricci tensor and $ R $ the associated Ricci scalar (see also \cite{Garcia2004} for a review on the Cotton tensor
and associated objects derived from it). One then calculates 
its matrix norm,  $ \left | H_{ij} \right | $, i.e. the square root of the largest eigenvalue of $ C_{ij} C^j_{\; k} $ and normalizes it by a scalar constructed from the covariant derivative of $ R_{ij} $, to provide a local measure of the accuracy of IWM--CFC
\begin{equation}
H: = \frac{\left | H_{ij} \right |}{\sqrt{D_i R_{jk} D^i R^{jk}}} \label{eq:H_local} \; .
\end{equation}
Finally, the quantity $H$ is further normalized by considering the following baryonic density weighted norm 
\begin{equation}
\left\langle H \right\rangle_{\rho} = \frac{\int d^3 x H \sqrt{\gamma} \rho W}{\int d^3 x \sqrt{\gamma} \rho W} \label{eq:H_rho} \; ,
\end{equation}
as a measure of the deviation from conformal flatness. In (\ref{eq:H_rho}), $ \gamma $ is the determinant of the 3-metric $ \gamma_{ij} $ and $ W = 1 / \sqrt{1-\gamma_{ij} \upsilon^i \upsilon^j } $ is the Lorentz factor, with $ \upsilon^i $ the 3-velocity.

The construction of $ \left\langle H \right\rangle_{\rho} $ is not straightforward, since the Cotton--York tensor that is involved needs the computation of third-order derivatives of $ \gamma_{ij} $. A first step is to express the spatial part of metric (\ref{eq:KEH_metric}) as a function of the compactified radial coordinate $ s =r/(r+r_{\mathrm{e}})$ \cite{CST92} and of $ m=\cos \theta $, which are the grid coordinates in {\tt rns}. Evaluating the Cotton--York tensor (\ref{eq:CY_tensor}) in the case of axisymmetry, yields the simple form
\begin{equation}
C_{ij}=\left( 
\begin{array}{ccc}
0 & 0 & C_{s \phi} \\
0 & 0 & C_{m \phi} \\
C_{s \phi} & C_{m \phi} & 0
\end{array}
\right) \; , \label{eq:CY_axisym}
\end{equation}
where the components $ C_{s \phi} $ and $ C_{m \phi} $ involve first-, second- and third-order partial derivatives of the metric functions\footnote{These calculations were performed with the help of a computer algebra program and the numerical results obtained were cross-checked with those of {\tt rns} by (i) specifying different polynomial functions as metric potentials, (ii) performing the same calculation in {\tt rns} and in the computer algebra program and (iii) ensuring that the results from both sources are in agreement.}.
Their detailed expressions are
\begin{align}
C_{s \phi} =
& B_{\mathrm{c}}
\left\{
\left[ 6 s (1-s) - \frac{m^2 + 1}{m^2 - 1} \right] \left( \mu - \bar{\mu} \right)_{,s} - s (s-1) (2s - 1) \left[ 3 \left( \mu - \bar{\mu} \right)_{,ss}  
\right. \right.
\nonumber \\
& \left. \left.
- 2 \left[ \left( \mu - \bar{\mu} \right)_{,s} \right] ^2\right] - s^2 \left( s-1 \right)^2 \left[ \left( \mu - \bar{\mu} \right)_{,sss} - 2 \left( \mu - \bar{\mu} \right)_{,s} \left( \mu - \bar{\mu} \right)_{,ss} \right]
\right. 
\nonumber \\    
& + m \left[ \left( \mu - \bar{\mu} \right)_{,sm} - 2 \left( \mu - \bar{\mu} \right)_{,s} \left( \mu - \bar{\mu} \right)_{,m} \right] + \left(m^2 -1 \right) \left[ \left( \mu - \bar{\mu} \right)_{,smm} 
\right.
\nonumber \\ 
& \left. \left.  
 - 2 \left( \mu - \bar{\mu} \right)_{,s} \left( \mu - \bar{\mu} \right)_{,mm} \right]
\right\} \; ,\label{eq:C_sphi_component}
\end{align}
and
\begin{align}
C_{m \phi}  =
& B_{\mathrm{c}}
\left\{
- 3 \left( \mu - \bar{\mu} \right)_{,m} - m \left[ 5  \left( \mu - \bar{\mu} \right)_{,mm} - 2  \left[ \left( \mu - \bar{\mu} \right)_{,m} \right]^2 \right]- \left(m^2 -1 \right) 
\right.
\nonumber \\
&  \times \left[  \left( \mu - \bar{\mu} \right)_{,mmm}
- 2  \left( \mu - \bar{\mu} \right)_{,m}  \left( \mu - \bar{\mu} \right)_{,mm} \right] +  s (s-1) (2 s -1) \left[  \left( \mu - \bar{\mu} \right)_{,sm}  
\right. 
\nonumber \\
& \left.  
- 2  \left( \mu - \bar{\mu} \right)_{,s} \left( \mu - \bar{\mu} \right)_{,m} + \frac{2 m}{m^2 -1}  \left( \mu - \bar{\mu} \right)_{,s} \right] + s^2 \left( s-1 \right)^2 \left[ \left( \mu - \bar{\mu} \right)_{,ssm}   
\right.
\nonumber \\
& \left. \left.
- 2 \left( \mu - \bar{\mu} \right)_{,ss} \left( \mu - \bar{\mu} \right)_{,m} + \frac{2 m}{m^2 -1} \left( \mu - \bar{\mu} \right)_{,ss} \right]
\right\} \; . \label{eq:C_muphi_component}
\end{align}
In the previous expressions, $ B_{\mathrm{c}} $ is a common multiplicative factor given by
\begin{equation}
B_{\mathrm{c}} = 
\frac{r_{\mathrm{e}} \; e^{(\gamma-\rho)/2}}{2 (s-1)^2} 
\; .\label{eq:CY_coef}
\end{equation}
Furthermore, we note that since $ C_{ij} $ is symmetric, its largest eigenvalue is the matrix norm $ \left | H_{ij} \right | $ involved in (\ref{eq:H_local}). 
Because $ C_{ij} $ vanishes in IWM--CFC, the idea put forward in \cite{Miller_Gressman_Suen} is to use a solution in full GR and assume that $ \left\langle H \right\rangle_{\rho} $ is representative of its deviation from conformal flatness. In \cite{Miller_Gressman_Suen} it was applied for neutron star binary inspirals. This has also been used in \cite{ott07_CFC_acc} for studying the rotating collapse of stellar iron cores.

\section{Results}
\label{sec:results}
\begin{figure}
\centering
\subfloat[]{
    \includegraphics[width=94mm]{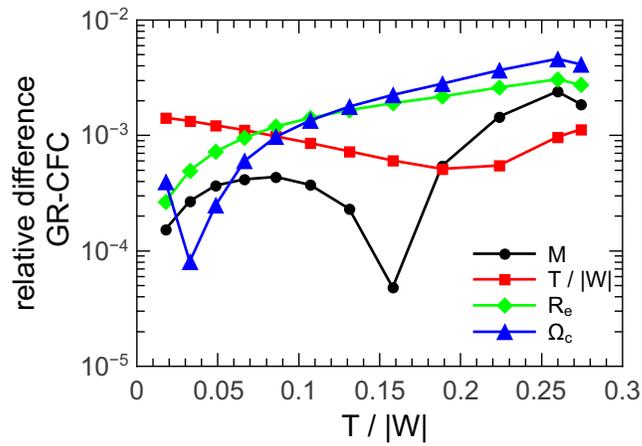}
        \label{fig:seqA_4physquant_log}
        }
\\
\subfloat[]{            
        \includegraphics[width=94mm]{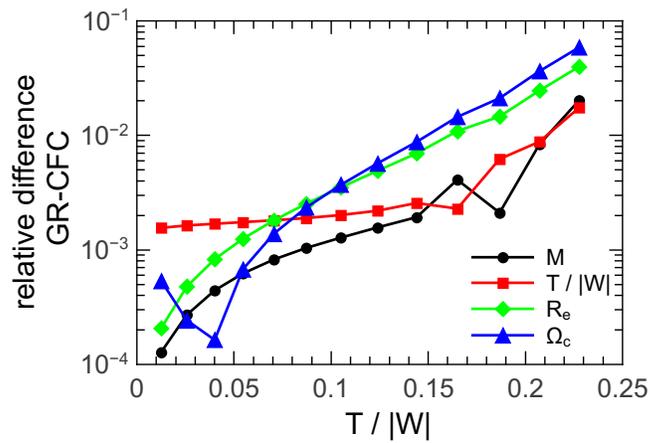}
        \label{fig:seqB_4physquant_log}
        }
\\
\subfloat[]{            
        \includegraphics[width=94mm]{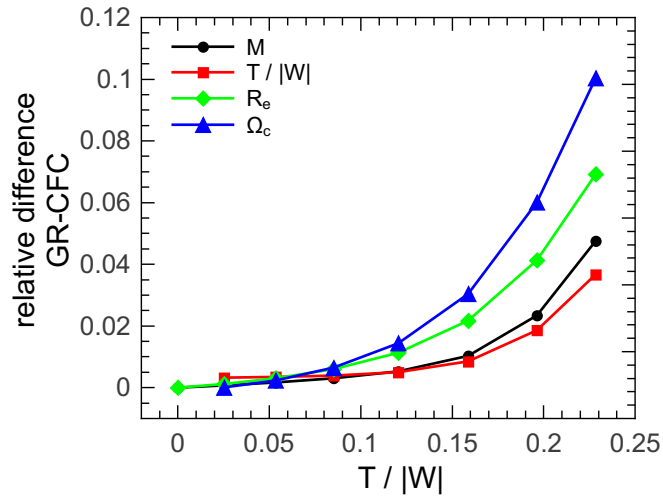}
        \label{fig:seqC_4physquant_linear}
        }               
\caption{Absolute values of the relative difference between full GR and IWM--CFC approximation for the gravitational mass $ M $, the ratio of rotational to gravitational binding energy $ T / |W| $, the circumferential radius $R_e$ at the equator and the angular velocity at the center of the configuration $\Omega_c$. \protect\subref{fig:seqA_4physquant_log} sequence A, \protect\subref{fig:seqB_4physquant_log} sequence B, \protect\subref{fig:seqC_4physquant_linear} sequence C.}
\label{fig:4physquant_log}
\end{figure}

\begin{figure}
\centering
\subfloat[]{
        \includegraphics[width=94mm]{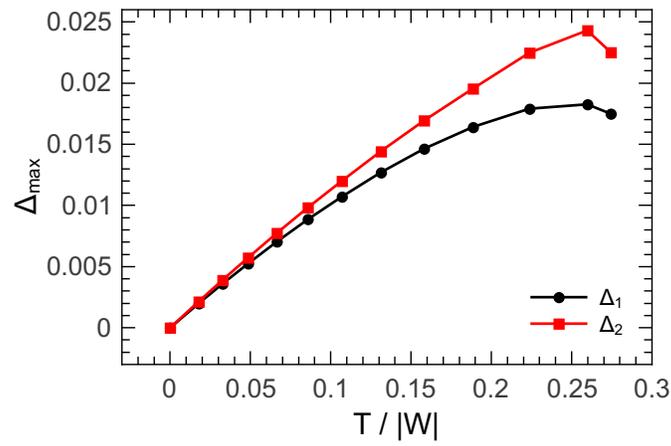}
        \label{fig:seqA_Delta_max}
        }
\\              
\subfloat[]{
        \includegraphics[width=94mm]{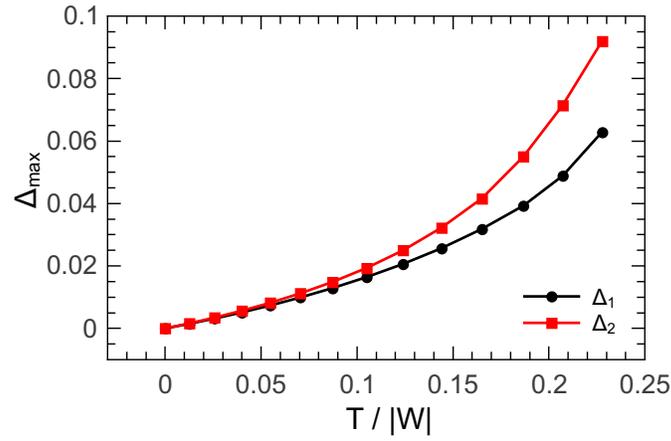}
        \label{fig:seqB_Delta_max}
        }
\\              
\subfloat[]{
        \includegraphics[width=94mm]{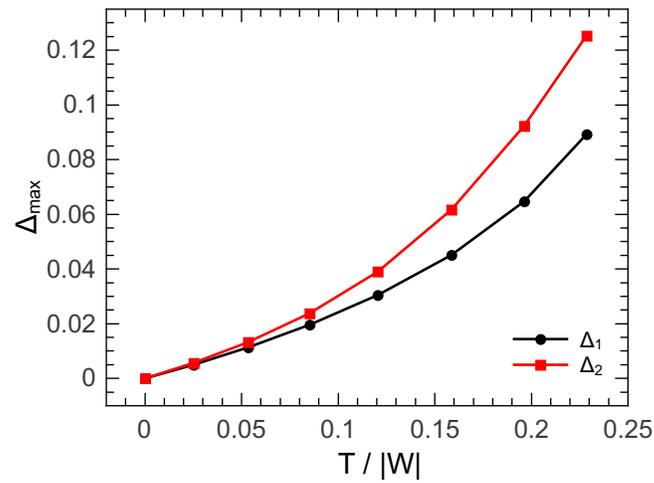}
        \label{fig:seqC_Delta_max}
        }        
\caption{Diagnostics $\Delta_1$ and $\Delta_2$ at the equator for all constructed models. \protect\subref{fig:seqA_Delta_max} sequence A, \protect\subref{fig:seqB_Delta_max} sequence B, \protect\subref{fig:seqC_Delta_max} sequence C.}
\label{fig:Delta_max}
\end{figure}

\begin{figure}
\centering
\subfloat[]{
        \includegraphics[width=94mm]{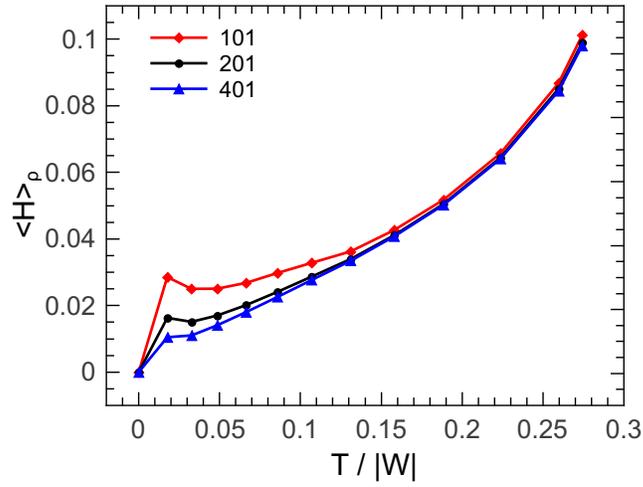}
        \label{fig:seqA_MGS}
        }
\\      
\subfloat[]{
        \includegraphics[width=94mm]{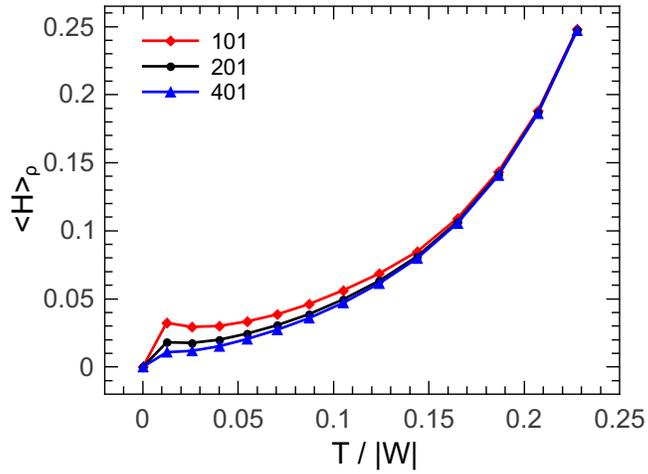}
        \label{fig:seqB_MGS}
        }
\\      
\subfloat[]{
        \includegraphics[width=94mm]{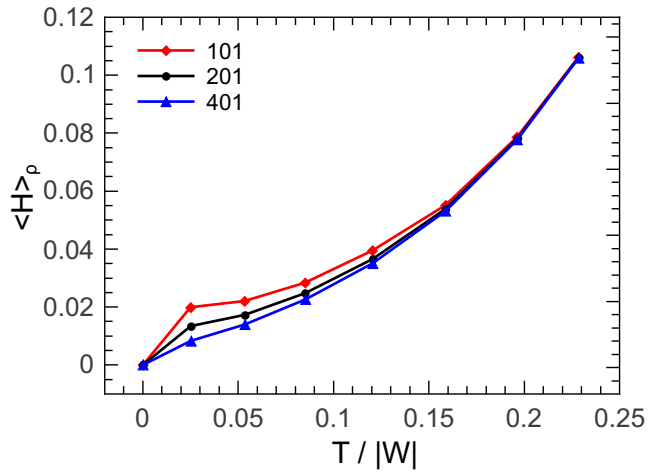}
        \label{fig:seqC_MGS}
        }        
\caption{Diagnostic $\left\langle H \right\rangle_{\rho}$ for all constructed models at different angular resolutions MDIV of the {\tt rns} code. Resolutions of 101, 201 and 401 points are represented as diamond, circle and triangle data points respectively. \protect\subref{fig:seqA_MGS} sequence A, \protect\subref{fig:seqB_MGS} sequence B, \protect\subref{fig:seqC_MGS} sequence C.}
\label{fig:MGS_measure}
\end{figure}

\begin{figure}
\centering
\includegraphics[width=0.8\columnwidth]{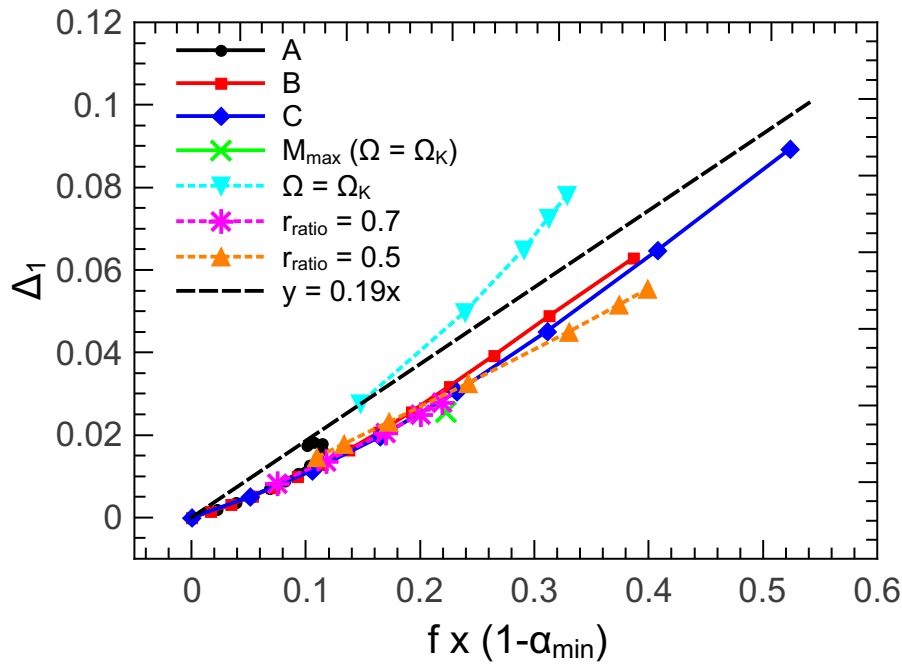}
\caption{Diagnostic $\Delta_1$ versus the flattening parameter $ f $ times $ (1-\alpha_{\rm{min}}) $. Solid lines represent polytropes with $ N=1 $ and $ K=100 $, while dotted lines represent polytropes with $ N=0.5 $ and $ K=1 $. Circle, rectangle and diamond points represent sequences A, B and C respectively. The maximum mass model of the corresponding Keplerian sequence is denoted with a cross. Down triangle points represent the Keplerian sequence of polytropes with $ N=0.5 $ and $ K=1 $. Star and up triangle points represent two differentially rotating sequences of fixed $ r_p / r_e $ with values of 0.7 and 0.5. The dashed line is our linear, empirical relation that approximates the data.}
\label{fig:Delta_trend}
\end{figure}

Figures \ref{fig:4physquant_log}, \ref{fig:Delta_max}, \ref{fig:MGS_measure} and \ref{fig:Delta_trend} summarize our findings. Figure \ref{fig:4physquant_log} shows the absolute value of the relative difference between full GR and IWM--CFC for four representative physical quantities, as a function of $ T / |W|, $ along the three different sequences of equilibrium models. We observe that along sequence A all relative differences saturate well below the 1\% level, which is explained by the fact that higher rotation leads to smaller central densities, which counteracts the effect of larger oblateness of the star. Along sequence B, the central density is fixed and the error increases practically monotonously, reaching up to 4\% for $R_e$ and 6\% for $\Omega_c$. Similar behaviour is encountered along sequence C, where maximum errors for $ M $ and $ \Omega_c $ reach 5\% and 10\% respectively.

Figure \ref{fig:Delta_max} displays the diagnostics $ \Delta_1 $ and $ \Delta_2 $ as a function of $ T / |W| $ along sequences A, B and C. For sequence A, $ \Delta_1 $ is less than 2\%, for sequence B it is around 6\%, and for sequence C it is below 9\%. The values of $\Delta_2$ rise higher as rotation increases but remain below 2.5\% for sequence A and around 9\% for sequence B, whereas for sequence C they are just over 12\%. The two measures are very similar for low and moderate rotation rates ($r_{\mathrm{p}}/r_{\mathrm{e}} < 0.5$). The maximum deviation for each model appears at the equator and at a slightly different value of the compactified radial coordinate $ s =r/(r+r_{\mathrm{e}}),$ as can be seen in Table \ref{tab:Delta_max_loc}.

As far as the diagnostic $\left\langle H \right\rangle_{\rho}$ is concerned, we note that it attains much larger numerical values along the constant-rest-mass sequence A than along the constant-central-density sequence B, if the original definition of \cite{Miller_Gressman_Suen} is followed (namely eq. (\ref{eq:H_local})), as described in Section \ref{sec:diagnostics}. This is in contrast to the other diagnostics displayed in Figures \ref{fig:4physquant_log} and \ref{fig:Delta_max}. Upon examination of the numerical evaluations of (\ref{eq:H_local}) and (\ref{eq:H_rho}), we find that this behaviour of $\left\langle H \right\rangle_{\rho}$ is attributed to the normalization of $ H $ by $ \sqrt{D_i R_{jk} D^i R^{jk}} $. As the central density decreases along sequence A, this quantity acquires much smaller values than at high densities (essentially, because one starts approaching the Newtonian limit). Therefore, the behaviour of $\left\langle H \right\rangle_{\rho}$ is not characterized by the numerator in (\ref{eq:H_local}), but by its denominator and we conclude that \textit{this choice of normalization is not appropriate for low-density configurations}. 

In contrast, along sequence B (which has a fixed central density) the numerical values of $\left\langle H \right\rangle_{\rho}$ do not exhibit similar behaviour and are only a factor of two larger than the diagnostic $ \Delta_2 $. Therefore, in order to overcome this obstacle, we have employed an alternative normalization for $ H $
\begin{equation}
H = \frac{\left | H_{ij} \right |}{\sqrt{ \left( D_i R_{jk} D^i R^{jk} \right)_0}} \label{eq:H_local_nonrot} \; ,
\end{equation}
where the denominator is always evaluated for the non-rotating model of the sequences. Apart from the normalization, $\left\langle H \right\rangle_{\rho}$ is computed as described in Section \ref{sec:diagnostics}. 

Figure \ref{fig:MGS_measure} displays the (modified) diagnostic $\left\langle H \right\rangle_{\rho}$ constructed from the Cotton--York tensor, along the three equilibrium sequences A, B and C. In all cases, the value of $\left\langle H \right\rangle_{\rho}$ increases significantly, as the rotation rate increases, so that the qualitative behaviour of this diagnostic is as expected. We notice that while for sequences A and B, $\left\langle H \right\rangle_{\rho}$ has larger values than $ \Delta_1 $ and $ \Delta_2 $, for sequence C its values are comparable to those of $ \Delta_1 $ and $ \Delta_2 $. This is attributed to the fact that the denominator of (\ref{eq:H_local_nonrot}) for sequence C is larger than for sequences A and B, as sequence C has a greater central density than both sequences A and B. We found that the numerator of (\ref{eq:H_local_nonrot}) is more dependent to the degree of rotation and therefore since all sequences have a comparable range of $r_p / r_e$ ratios, it is not responsible for the different behaviour of $\left\langle H \right\rangle_{\rho}$ between the three sequences when compared to the behaviour of the diagnostics $ \Delta_1 $ and $ \Delta_2 $.

Figure \ref{fig:MGS_measure} also shows a convergence study of $\left\langle H \right\rangle_{\rho}$ at three different angular resolutions, showing satisfactory convergence. The radial resolution in all cases was fixed at 201 grid points. Notice that the equilibrium equations solved in {\tt rns} only involve up to second-order derivatives, while the numerical evaluation of the Cotton--York tensor requires third-order derivatives. Because the numerical grid is not adapted to the surface of the star, Gibbs phenomena are inevitable. These do not pose a significant problem for derivatives up to second order, but when evaluating third-order derivatives the Gibbs phenomena do not allow for improved convergence past roughly 201 radial grid points. Using a surface-adapted grid would cure this problem. In order to subdue this effect and have as smooth third-order derivatives as possible, we doubled the grid spacing for third-order partial derivatives, instead of using a standard central 
difference formula.
This technique was already used in  \cite{stergfried95} to suppress numerical point-to-point noise in second-order derivatives and we extend it here to third-order derivatives. 

Figure \ref{fig:Delta_trend} shows a trend that emerges if one plots the error indicator $ \Delta_1 $ versus the product of the flattening parameter $ f $
\begin{equation}
f := 1-\frac{r_p}{r_e} \label{eq:flattening} \; ,
\end{equation}
times $ (1-\alpha_{\rm min}) $, where $ \alpha_{\rm min} $ is the minimum of the lapse function (\ref{eq:lapse_shift_psi}). Sequences A, B and C are shown together with the maximum mass model of the Keplerian sequence for the EOS with $ N=1 $ and $ K=100 $. Furthermore, we construct two differentially rotating sequences with $ r_p/r_e $ values of 0.7 and 0.5 for the EOS with $ N=0.5 $ and $ K=1 $. All the previously described data points fall roughly along a line and an empirical formula that describes well the numerical data is the straight line through the origin, of slope 0.15. For the EOS with $ N=0.5 $ and $ K=1 $, we also show the (uniformly rotating) Keplerian sequence, which lies just above all the other data points and is described well by a straight line through the origin, of slope 0.23. Taking into account the range of polytropic EOSs between $ N=0.5$ and $ N=1.0 $, we find that the approximate,  empirical formula   
\begin{equation}
\Delta_1 = 0.19\times f \times(1-\alpha_{min})  \label{eq:Delta_f_lapse_trend} \; ,
\end{equation} 
can be used to obtain a good estimate for the error indicator $ \Delta_1 $ from the flatness parameter and the minimum value of the lapse of a given model.
We notice that the error of the above empirical formula falls within the requirements of the present study (we are mainly interested in  the order of magnitude of the error when assuming the IWM--CFC approximation). For the same reason, we only use the error indicator $\Delta_1$, since $\Delta_2$ attains values that are not significantly different.

\begin{table}
\centering
\caption{\label{tab:Delta_max_loc}Diagnostics $\Delta_1$ and $ \Delta_2 $ and  compactified radial coordinate $s$, where they attained their maximum value. Detailed characteristics of the models can be found in Table 1 of \cite{SAF2004}.}
\begin{tabular}{@{}llllll}
\hline\noalign{\smallskip}
Model & $ r_p / r_e $ & $ T /|W| $ & $s$ at max& $\Delta_1$ & $\Delta_2$ \\
    & & ($ \times 10^{-1}$) & & ($ \times 10^{-2}$) & ($ \times 10^{-2}$) \\
\noalign{\smallskip}\hline\noalign{\smallskip}
A0 & 1.000 & 0.000 & -- & 0.0 & 0.0\\
A1 & 0.930 & 0.177 & 0.42 & 0.2 & 0.2\\
A2 & 0.875 & 0.326 & 0.42 & 0.4 & 0.4\\
A3 & 0.820 & 0.485 & 0.42 & 0.5 & 0.6\\
A4 & 0.762 & 0.664 & 0.42 & 0.7 & 0.8\\
A5 & 0.703 & 0.858 & 0.42 & 0.9 & 1.0\\
A6 & 0.643 & 1.069 & 0.42 & 1.1 & 1.2\\
A7 & 0.579 & 1.311 & 0.42 & 1.3 & 1.4\\
A8 & 0.513 & 1.580 & 0.42 & 1.5 & 1.7\\
A9 & 0.444 & 1.884 & 0.42 & 1.6 & 2.0\\
A10 & 0.370 & 2.236 & 0.42 & 1.8 & 2.2\\
A11 & 0.294 & 2.597 & 0.42 & 1.8 & 2.4\\
A12 & 0.250 & 2.743 & 0.42 & 1.7 & 2.3\\
\\
B0 & 1.000 & 0.000 & -- & 0.0 & 0.0\\
B1 & 0.950 & 0.125 & 0.43 & 0.1 & 0.2\\
B2 & 0.900 & 0.257 & 0.42 & 0.3 & 0.3\\
B3 & 0.849 & 0.400 & 0.42 & 0.5 & 0.6\\
B4 & 0.800 & 0.546 & 0.42 & 0.7 & 0.8\\
B5 & 0.750 & 0.704 & 0.42 & 1.0 & 1.1\\
B6 & 0.700 & 0.872 & 0.42 & 1.3 & 1.5\\
B7 & 0.650 & 1.050 & 0.42 & 1.6 & 1.9\\
B8 & 0.600 & 1.239 & 0.42 & 2.1 & 2.5\\
B9 & 0.550 & 1.440 & 0.42 & 2.6 & 3.2\\
B10 & 0.500 & 1.650 & 0.42 & 3.2 & 4.2\\
B11 & 0.450 & 1.867 & 0.42 & 4.0 & 5.5\\
B12 & 0.400 & 2.072 & 0.43 & 4.9 & 7.1\\
B13 & 0.340 & 2.277 & 0.44 & 6.3 & 9.2\\
\\
C0 & 1.000 & 0.000 & -- & 0.0 & 0.0\\  
C1 & 0.900 & 0.251 & 0.42 & 0.5 & 0.6\\
C2 & 0.800 & 0.534 & 0.42 & 1.1 & 1.3\\
C3 & 0.700 & 0.851 & 0.42 & 2.0 & 2.4\\
C4 & 0.600 & 1.205 & 0.42 & 3.0 & 3.9\\
C5 & 0.500 & 1.587 & 0.42 & 4.5 & 6.2\\
C6 & 0.400 & 1.963 & 0.43 & 6.5 & 9.2\\
C7 & 0.300 & 2.285 & 0.43 & 8.9 & 12.5\\
\noalign{\smallskip}\hline
\end{tabular}
\end{table}

\section{Discussion}
\label{sec:discussion}
In order to answer the question which diagnostic performs better as a measure of the error of the IWM--CFC approximation, we should look at how well they correlate with the maximum errors encountered in physical quantities. The constructed measures $\Delta_1$ and $\Delta_2$ qualify as adequate choices. Their values and qualitative behaviour are in agreement with that of physical quantities $\Omega_c$ and $R_e$ implying that they are more sensitive to the degree of rotation than integrated physical quantities, such as $M$ and $T / |W|$. Preliminary results presented in \cite{IosifStergioulasNEB15} showed that the relative error $ \Delta_1 $ peaks around $s\simeq 0.4$, or $ r \simeq \frac{2}{3} r_{\mathrm{e}} $. Our current analysis confirms that the maximum deviation appears at around $ s \simeq 0.42-0.44 $ for all models considered in this survey. Therefore, if one aims to assign a single number to each model in order to evaluate the accuracy of IWM--CFC, then calculating $ \Delta_1 $ at the point where it attains its maximum value provides a simple way to do so for the case of isolated rotating stars.

The integrated quantity $ \left\langle H \right\rangle_{\rho} $ comes at second place behind $ \Delta_1 $ as a measure of the accuracy of IWM--CFC. Despite the fact that it is based on the Cotton--York tensor, which has the advantage of vanishing identically in IWM--CFC, overall its values do not correlate well with the relative errors in physical quantities. In addition, it demands the calculation of third order derivatives of the spatial metric tensor, while $ \Delta_1 $ is simpler to evaluate.

We note that $ \left\langle H \right\rangle_{\rho} $ works better as a diagnostic only if the modification (\ref{eq:H_local_nonrot}) is made. Otherwise, in its original form using (\ref{eq:H_local}) instead of (\ref{eq:H_local_nonrot}), it is unsuitable for constant-rest-mass sequences, that also include low-density models, such as sequence A. To put it differently, the numerator of (\ref{eq:H_local}) compares IWM--CFC to GR, while its denominator compares Newtonian theory to GR. Even though each one on its own may be a small number, e.g. of the order of $ \sim10^{-2} $, in the case of sequence A their ratio can reach considerably higher values, thus making this diagnostic unfit for these cases.

We stress that the 5\% error in integrated quantities (10\% for local quantities) is seen only for the most relativistic and most rapidly rotating, stable equilibrium models. For models that are somewhat less relativistic and/or moderately rotating (such as realistic proto-neutron stars created in a core collapse of a massive star or some hypermassive differentially rotating stars created after a binary neutron star merger event, that are not too compact) we find that the IWM--CFC approximation results in errors at the 1\% level. Therefore, the IWM--CFC approximation is a robust method for studying a range of typical neutron star models with differential rotation. It will be interesting to extend this study to a larger set of realistic (microphysical) EOSs, in order to determine more precisely the performance of the approximation for different choices of EOS and mass.

\begin{acknowledgements}
We are grateful to Andi Bauswein for useful discussions. This work was supported by an IKY--DAAD exchange grant (IKYDA 2012) and a Virgo EGO Scientific Forum (VESF) fellowship for doctoral studies by the European Gravitational Observatory (EGO-DIR-126-2012 / EGO-DIR-80-2013). Partial support also provided by ``NewCompStar", COST Action MP1304.
\end{acknowledgements}

\appendix

\section*{Appendix}
\label{sec:appendix}
\begin{figure}
\centering
\subfloat[]{
        \includegraphics[width=0.55\columnwidth]{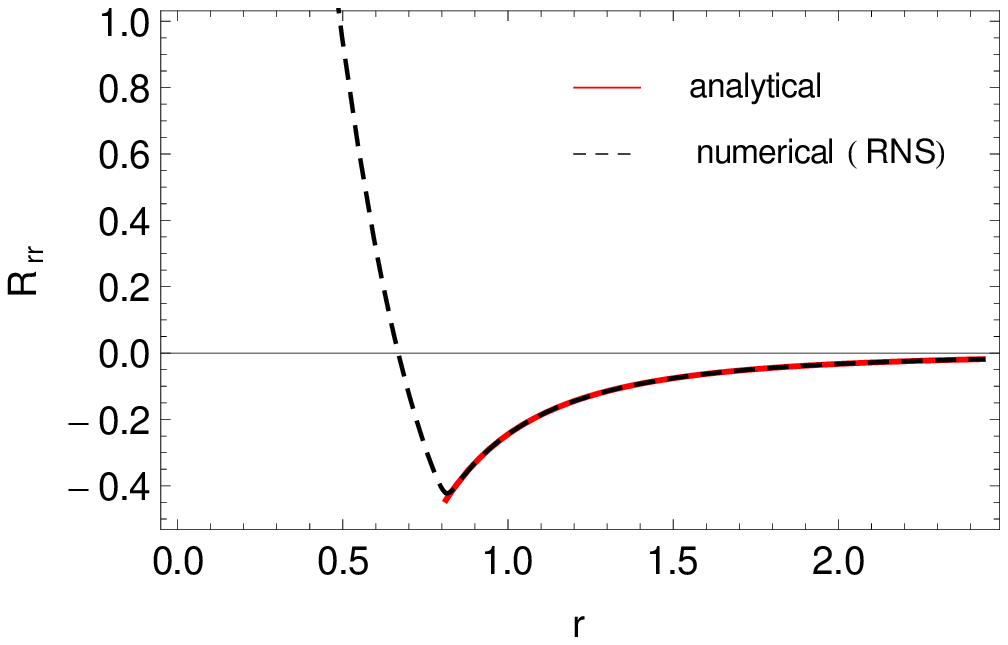}
        \label{fig:R_rr_eq}
        }
\\              
\subfloat[]{
        \includegraphics[width=0.55\columnwidth]{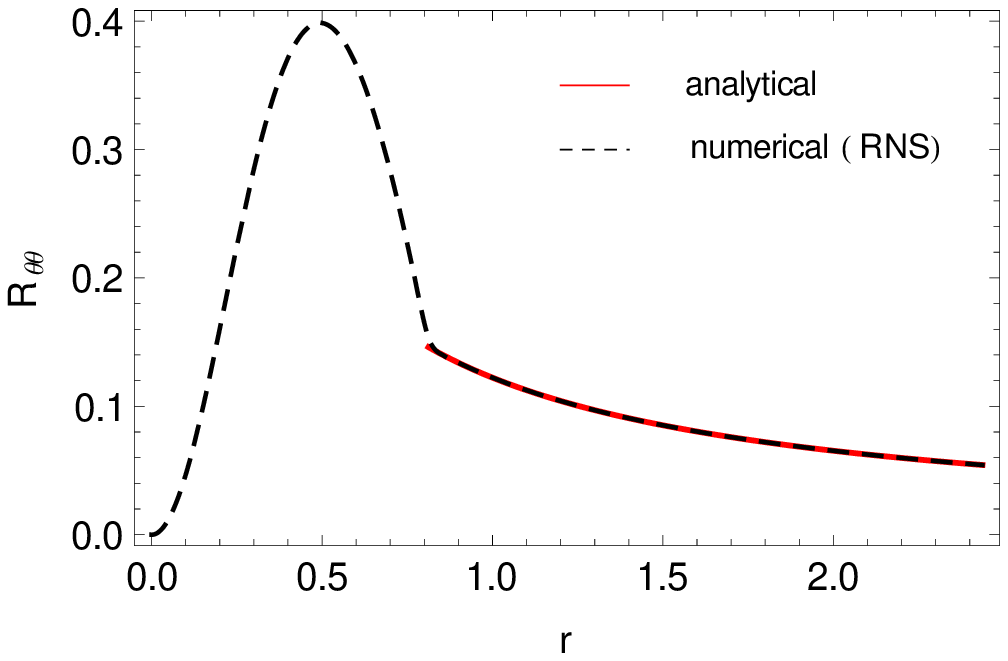}
        \label{fig:R_thth_eq}
        }
\caption{3-dimensional Ricci tensor components in isotropic Schwarzschild coordinates at the equator up to a distance of  $ 3 r_e $ ($ r_e = 0.8124$).\protect\subref{fig:R_rr_eq} $R_{rr} $, \protect\subref{fig:R_thth_eq} $R_{\theta \theta}$}
\label{fig:Ricci_comps_iso_eq}
\end{figure}

As an additional check for our numerical code, we evaluate the three-dimensional Ricci tensor components in isotropic Schwarzschild coordinates and compare with the corresponding analytic expressions for the exterior of the star, which are (see e.g. \cite{NR_baumshapi})
\begin{equation}
\label{eq:Ricci_comps_iso}
R_{rr} = - \frac{8 r M}{\left( 2 r^2 + M r \right)^2} \; , \qquad
R_{\theta \theta} = \frac{4 r^3 M}{\left( 2 r^2 + M r \right)^2} \; , \qquad 
R_{\phi \phi} = \sin ^2 \theta R_{\theta \theta} \; ,
\end{equation}
where $ M $ is the gravitational mass of the star. Notice that while in vacuum the 4-dimensional Ricci tensor vanishes, this is not the case in three dimensions. Figure \ref{fig:Ricci_comps_iso_eq} shows the numerical and analytic evaluation of the Ricci components at the equator for the nonrotating model of the sequences. The $ R_{\phi \phi} $ component is not shown, since it coincides with $ R_{\theta \theta} $. 

\bibliographystyle{spphys}                       
\bibliography{2014-IS}   

\end{document}